\documentclass[aps,pra,superscriptaddress,unsortedaddress, twocolumn]{revtex4}
\usepackage{amsmath, amssymb, mathtools}
\usepackage{xcolor, bm, soul}
\usepackage{comment}
\usepackage{float}
\usepackage{upgreek}
\usepackage{xfrac}
\usepackage{epstopdf,comment}
\usepackage{bbold}
\usepackage{url}
\usepackage[braket, qm]{qcircuit}
\usepackage{braket}
\usepackage[bookmarks=true,
            bookmarksnumbered=false,
            bookmarksopen=true,
            breaklinks=true,
            pdfborder={0 0 0},
            final=true,
            backref=none,
            colorlinks=true,
            linkcolor=blue,
            menucolor=blue,
            citecolor=blue,
            urlcolor=blue]{hyperref}
\usepackage{ulem}
\usepackage{caption}

\begin{document}

\author{Tatiana A. Bespalova}
\affiliation{Universit\'e de Strasbourg and CNRS, ISIS (UMR 7006), 67000 Strasbourg, France}

\author{Karlo Deli\'c}
\affiliation{Department of Mathematics, Faculty of Science, University of Zagreb, Bijeni\v{c}ka cesta 30, 10000 Zagreb, Croatia}

\author{Guido Pupillo}
\affiliation{Universit\'e de Strasbourg and CNRS, ISIS (UMR 7006), 67000 Strasbourg, France}

\author{Francesco Tacchino}
\affiliation{IBM Quantum, IBM Research - Zurich, S\"aumerstrasse 4, 8803 R\"uschlikon, Switzerland}

\author{Ivano Tavernelli}
\affiliation{IBM Quantum, IBM Research - Zurich, S\"aumerstrasse 4, 8803 R\"uschlikon, Switzerland}

\date{\today}

\title{Simulating the Fermi-Hubbard model with long-range hopping on a quantum computer}

\begin{abstract}
We investigate the performance and accuracy of digital quantum algorithms for the study of static and dynamic properties of the fermionic Hubbard model at half-filling with next-nearest neighbour hopping terms. We provide quantum circuits to perform ground and excited states calculations, via the Variational Quantum Eigensolver (VQE) and the Quantum Equation of Motion (qEOM) approach respectively, as well as product formulas decompositions for time evolution. We benchmark our approach on a chain with $L=6$ sites and periodic boundary conditions, computing the charge and spin gaps, the spectral function and spin-spin dynamic correlations. Our results for the ground state phase diagram are in qualitative agreement with known results in the thermodynamic limit. Finally, we provide concrete scalings for the number of gates needed to implement our protocols on a qubit register with all-to-all connectivities and on a heavy hexagonal coupling map.
\end{abstract}

\maketitle

\section{Introduction}

The Fermi-Hubbard Model (FHM) is widely used to describe systems of interacting electrons in lattice structures, including crystals~\cite{arovas2022hubbard}. 
The FHM is particularly relevant in understanding the properties of strongly correlated materials, e.g., high-temperature superconductors and several classes of magnetic materials. While the basic FMH, featuring only nearest neighbours hoppings, already captures many important properties, it is well known that a comprehensive, quantitative description of electron behaviour in materials can only be obtained by including couplings beyond nearest neighbours (e.g., with next-to-nearest neighbour hopping terms). Interestingly, interactions beyond nearest neighbours can significantly impact the model properties, giving rise to rich phase diagrams, as, for instance, the Mott-Hubbard transition at half-filling. It was also shown that they are required to induce thermalization in the one-dimensional FHM~\cite{biebl2017}. 

While the FHM is simple to formulate, it is in practice challenging to solve numerically, due to its strongly correlated character~\cite{leblanc2015solutions, qin2022hubbard}. Classical solutions of the static Hubbard model are usually confined to two-dimensional lattices of about 10 $\times$ 10 sites, while the situation gets even more critical when dealing with dynamic properties. In this case, due to the steady increase in the level of entanglement over time, the classical simulation of time-dependent properties becomes very expensive already in one dimension. 

In recent years, quantum computing has emerged as a new promising approach to tackle simulation problems in the natural sciences, including e.g., physics~\cite{Feynman1982, liu2020, arute2020observation, Mazzola_rew_2023, Miessen2023, charles2024, Schuhmacher_MLHEP_2023, wu2023variational, Dimeglio2024quantum}, chemistry~\cite{Moll_rew_2018, cao2019quantum, Ollitrault_rew_2021, Motta2021, Mazzola_rew_2023, shang2023}, biology~\cite{emani2021,basu2023quantumenabled} and materials science~\cite{clinton2024,IBM_Mat_Rew_2024}, as well as in the social sciences, such as economics and finance~\cite{herman2023,abbas2023quantum}. 
While state-of-the-art quantum processors are still too noisy to allow a reliable implementation of many of the most advanced algorithms, several near-term alternatives have already been proposed and tested. Notably, these include proposed solutions for computing ground and excited states of model Hamiltonians~\cite{miessen2021quantum, clinton2024} and quantum chemistry systems~\cite{sokolov2020quantum, ollitrault2020quantum, feniou2023}, the simulation of quantum dynamics~\cite{Miessen2023, bultrini2023}, and the evaluation of many-body Green's functions~\cite{endo2020calculation, jamet2022quantum, gomes2023, LibbiGreen2022, RizzoGreen2022, selisko2024dynamic}. Importantly, recent experiments demonstrated that the regime of quantum utility, in which existing quantum computers can tackle problems for which only approximate classical solutions are available, is within reach~\cite{kim2023evidence}. This level of performance is made possible by steady hardware improvements (e.g., higher gate fidelities and faster gate operations) and by the use of suitable error mitigation techniques~\cite{van2023probabilistic,kim2023evidence,filippov2023scalable}.

In this work, we investigate the performance and accuracy of quantum algorithms for the study of the static and dynamic properties of the FHM model with hopping terms extended to next-nearest neighbours on a linear lattice with periodic boundary conditions. 
We first focus on the design and implementation of a quantum variational ansatz for the study of ground state properties. We evaluate charge and spin gaps as functions of the ratio between nearest and next-to-nearest hopping terms, observing signatures of a transition between a gapless and a gapped phase.  In~\cite{daul2000phase, torio2003phase, tocchio2010interaction} it was shown that the half-filled Hubbard chain with next-nearest-neighbour hopping exhibits commensurate and incommensurate disordered magnetic insulating phases, a spin-gapped metallic phase, and a frustrated antiferromagnetic insulator. In our work, we compare these results with our quantum computing simulations of a periodic 6-sites Hubbard model to assess the quality of the proposed quantum algorithms, and the impact of the finite size effects. 

Furthermore, by means of variational and Green’s functions techniques, we also investigate excited states' properties of the same periodic one-dimensional chain.  We then analyze its time-dependent properties through the implementation of a time-evolution algorithm based on product formulas. Finally, after assessing the quality of our approaches by means of exact numerical calculations, we perform a thorough resource estimation, in terms of number of qubits and quantum gate operations, to inform future implementations of the proposed algorithms on quantum processors. 

\section{Description of the Model} \label{Model}
\begin{figure}
    \centering
    \includegraphics[width=0.9\linewidth]{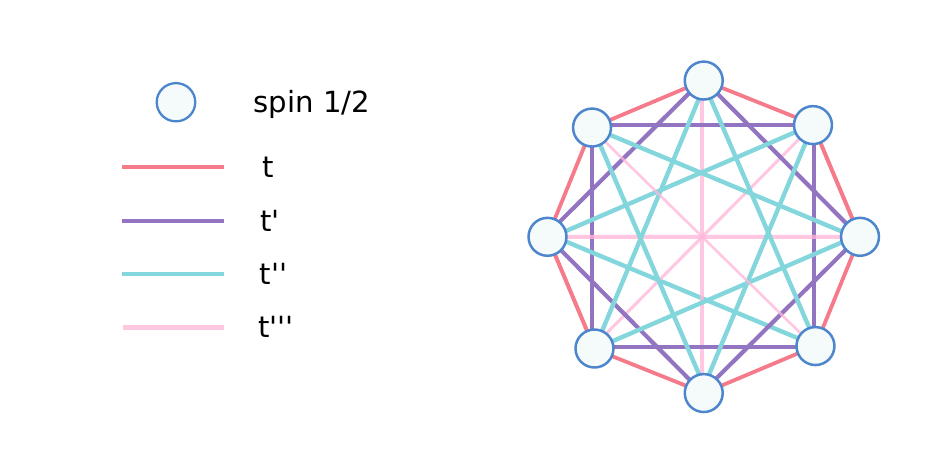}
    \caption{Sketch of the connectivity of a periodic one-dimensional Hubbard chain with hopping terms beyond nearest neighbours. $U$ is the onsite interaction, while $t$, $t'$, $t''$, $t'''$ are hopping coefficients.}
    \label{fig:HubbardModelSketch}
\end{figure}

We focus on the 1D Fermi-Hubbard model beyond nearest neighbour interactions with periodic boundaries (Fig.~\ref{fig:HubbardModelSketch}). The hopping coefficients between two sites depend only on the distance along the sequence, while the repulsive term applies only to opposite spins occupying the same site. 

The Hamiltonian can be written as
\begin{equation}
\begin{split} \label{eq:Hamiltonian}
    \mathbf H =-&\sum\limits_{\sigma, j=1}^{{L}}\left[
    t \, \mathbf c_{j,\sigma}^\dagger \mathbf c_{j + 1, \sigma}  +
    t' \, \mathbf c_{j,\sigma}^\dagger \mathbf c_{j + 2, \sigma} + \right. \\  &\left. t'' \, \mathbf c_{j,\sigma}^\dagger \mathbf c_{j + 3, \sigma} +
    \text{\dots} + \text{H.c.} \right] +U\sum\limits_{j = 1}^{{L}} \mathbf n_{j, \uparrow} \mathbf n_{j, \downarrow} \text,
\end{split}
\end{equation}
where $\mathbf c_{j,\sigma}^\dagger$ describes the creation operator for a fermionic particle with spin $\sigma$ at site $j$ of a chain of length ${{L}}$, and $\mathbf n_{j, \sigma} = \mathbf c_{j,\sigma}^\dagger \mathbf c_{j,\sigma}$. Periodic boundary conditions are enforced by setting $j + {{L}} \coloneqq j$. The last term in the Hamiltonian describes onsite interaction of strength $U$, while the terms inside square brackets describe hopping terms between first, second, third, \ldots  nearest neighbours  sites with hopping coefficients $t$, $t'$, $t''$, \ldots, respectively. In Eq.~\eqref{eq:Hamiltonian} and throughout the paper we use $\hbar=1$.

In the following, we will restrict our analysis to positive hopping coefficients and onsite interaction term $U$, and we will consider systems with $L=6$ lattice sites with hopping terms between nearest and next-to-nearest sites. The value of the nearest-neighbours hopping energy $t$ is used as our energy unit.  A generalization to the case with negative onsite interaction can be easily obtained by applying the Shiba map~\cite{shiba1972} to the Hamiltonian.

The properties of the FHM in the thermodynamic limit have been extensively studied and can exhibit a variety of interesting phenomena, e.g., a Mott insulating behaviour and an anti-ferromagnetic phase. In this work, we focus on the half-filled case with no spin imbalance and compare our results to previous studies~\cite{daul2000phase, torio2003phase, tocchio2010interaction}. In the thermodynamic limit, the ground state of the $t$\--$t'$ Hubbard model at half-filling is predicted to be \textit{(I)} an insulator with gapless spin excitations (conventionally labelled as C0S1, where CnSm represents a phase with $n$ gapless charge modes and $m$ gapless spin modes~\cite{balents1996weak}) for $t'/t < 1/2$ and finite $U/t$~\cite{lieb1968absence}, \textit{(II)} a fully-gapped spontaneously dimerized insulator (C0S0) for $t'/t > 1/2$ and large $U/t$~\cite{daul2000phase, arita1998density}, and, \textit{(III)} a spin-gapped metal (C1S0) with strong superconducting fluctuations for $t'/t > 1/2$ and small $U/t$~\cite{fabrizio1996superconductivity}.

\section{Methods} \label{Methods}
In this section, we describe the quantum algorithms that we will employ in our study. The guiding principle for the selection of such algorithms is that to ensure, given their quantum circuit requirements, compatibility with near-term quantum architectures, particularly in the range of 50 to a few hundreds of qubits.
 
In Sec.~\ref{Methods.VQE} we describe the use of Variational Quantum Eigensolver (VQE) to prepare approximations of the FH ground state. The prepared ground state is further used in Sections~\ref{Results.Gaps},~\ref{Results.SpectralFunction} and~\ref{Results.Dynamics} to compute excited states of the system and to reconstruct the spectral function and dynamic correlations. Specifically, in Sec.~\ref{Methods.qEOM} we describe the Quantum Equation of Motion (qEOM) method that we use to compute the excited states of the system, while in Sec.~\ref{Results.SpectralFunction} we show the calculation of the spectral function in the Lehmann representation~\cite{lehmann1954} with the use of excited states obtained from qEOM. Finally, Sec.~\ref{Methods.Trotterization} gives a brief description of the Trotterisation procedure to compute the dynamics of the system.

\subsection{Variational quantum eigensolver} \label{Methods.VQE}
A first, fundamental task to address when studying the FHM is that of finding a good approximation of the ground state in a given particle number and total spin sector. For this, we employ the variational quantum eigensolver (VQE)~\cite{peruzzo2014variational, mcclean2016theory}, a hybrid quantum-classical algorithm that leverages both quantum registers, for preparing and evaluating trial quantum states, and classical subroutines, for optimization purposes. 

VQE was recently applied to find the ground state of the nearest neighbour Fermi-Hubbard model using state-of-the-art techniques~\cite{stanisic2022observing, mineh2022solving, anselme2022simulating, cade2020strategies}.
The core idea behind the VQE approach is to use a quantum computer to prepare a parametrized quantum state $\ket{\psi (\theta)}$ that, through suitable refinements, can approximate the target ground state. Following the variational principle, the parameters $\theta $ characterizing this state are varied to minimize the expectation value of the Hamiltonian $\mathbf H$, i.e., the energy. A classical optimizer is used to iteratively adjust the parameters $\theta$. 

The VQE algorithm consists of the following steps: (i) Ansatz construction, namely choosing a suitable parameterized quantum circuit $U(\mathbf\theta)$ to prepare a family of trial states $\ket{\psi (\theta)}= U(\mathbf\theta)\ket{0}$ from a fixed reference; (ii) parameter initialization, i.e., choosing initial values for $\mathbf\theta$ to start the optimization procedure; (iii) energy evaluation, i.e., measuring the energy as $\bra{\psi(\mathbf\theta)} \mathbf H \ket{\psi(\mathbf\theta)}$; (iv) parameter optimization. Steps (iii) and (iv) are repeated until convergence.
To construct a suitable Ansatz for the Hamiltonian in Eq.~\eqref{eq:Hamiltonian}, we will also take into account the relevant symmetries of the problem, thus constraining the search space for VQE in a physically meaningful way. The specific realisation of these constraints is described in Sec.~\ref{Results.GroundState}.
The mapping from Fermion to qubit space was achieved through a  Jordan-Wigner transformation~\cite{jordanwigner1928};  alternatives approaches could also be considered such as the Bravyi-Kitaev maps~\cite{bravyi2002fermionic, seeley2012bravyi, tranter2015b} or the more recent Bonsai protocol~\cite{miller2023bonsai} and ternary trees approaches in general~\cite{vlasov2019clifford, parella2024reducing, miller2024treespilation}. 

\subsection{Quantum Equation of Motion} \label{Methods.qEOM}
To compute the excited eigenenergies of Eq.~\eqref{eq:Hamiltonian}, we will use a slightly modified version of the quantum Equation of Motion (qEOM) approach introduced in Ref.~\cite{ollitrault2020quantum}. This method constructs an excited state of the system, $\ket{n}$, as a linear combination of suitable excitation operators applied on the ground state (or, more realistically, to its approximation obtained, e.g., in a previous VQE step). 

More specifically, the $n$-th excited state creation operator $\hat O^\dagger_n = \ket{n} \bra{0}$ is constructed as
\begin{equation}
    \hat O^\dagger_n \approx \sum_\mu X_\mu (n) \hat E_\mu\text, 
\end{equation}
where $\hat E_\mu$ are elementary excitation operators and $X_\mu (n)$ are complex variational parameters. This approximate relation is exact in the limit in which $\{E_{\mu}\}$ is a complete spanning set.  In practice, one often limits this set to only a subset of relevant excitations. In our case, for instance, we will only use those operators that preserve the half-filling property of the system. The exact choice of the excitation operators used in the expansion depends on the nature of the system, the targeted precision, and the level of excitation one would like to achieve. We will describe our choice of operators in Sec.~\ref{Results.SpectralFunction}.

To evaluate the variational parameters $X_\mu$ corresponding to the $n$-th eigenstate of the system, we express the energy of the $n$-th eigenstate $\ket{n}$ as

\begin{align}
    \notag
    E_{n} &= \frac{\bra{n} \mathbf H \ket{n}}{\braket{n|n}}= \frac{\bra{0} \hat O_n \mathbf H\hat O^\dagger_n\ket{0}}{\bra{0}\hat O_n  \hat O^\dagger_n\ket{0}} = \\ &=\frac{\sum_{\mu\mu'}X^*_\mu (n) X_{\mu'} (n)\bra{0} \hat E_\mu^\dagger \mathbf H \hat E_{\mu'}\ket{0}}{\sum_{\mu\mu'}X^*_\mu (n) X_{\mu'} (n)\bra{0} \hat E_\mu^\dagger \hat E_{\mu'} \ket{0}}\text.
    \label{eq:qEOM}
\end{align}

The variable variational parameters $X_{\mu}$ correspond to the extrema of Eq.~\eqref{eq:qEOM}. Thus, they correspond to the eigenvectors of the generalized eigenvalue problem
\begin{equation} \label{eq:GeneralizedEigenvalue}
    \bra{0} \hat E_\mu^\dagger \hat E_{\mu'} \ket{0} E_{n} X_\mu = \bra{0} \hat E_\mu^\dagger \mathbf H \hat E_{\mu'} \ket{0} X_\mu\text.
\end{equation}
In practical implementations, the matrix elements appearing in Eq.~\eqref{eq:GeneralizedEigenvalue} should be measured on a quantum device as expectation values on the (approximate) ground state, while the generalized eigenvalue problem itself is solved on a classical computer.

Thanks to its modest requirements in terms of quantum circuit complexity (which do not exceed those for the corresponding ground state calculations), the qEOM approach is particularly well suited for near-term quantum processors. However, it is also subject to certain limitations and numerical instabilities, which may lead to unphysical or unwanted states. These may include, for instance, states outside the subspace of interest (half-filling sector) or with wrong symmetries (e.g., spin configuration), which  may occur when the ground state produced by VQE is not fully respecting the imposed symmetries, or when the latter are broken by the effect of hardware noise. Degenerate solutions, corresponding to states that share the same energy, also require special attention.  In the first case, we can identify the unphysical states by computing the norm or the expectation value associated to the symmetry of interest, e.g., the total number of particles if we are interested in a particular particle occupation sector of the full Fock space. 

\subsection{Time propagation} \label{Methods.Trotterization}
To simulate the dynamics of the system in a digital quantum computing setup we will make use of a well-known systematic decomposition approach based on product formulas~\cite{miessen2021quantum}, and specifically the so-called Trotterization method~\cite{Lloyd1996}. This amounts to a factorization of the unitary time evolution operator $e^{-\imath\mathbf H\tau}$ into simpler ones and its application in discrete, small time steps $\Delta \tau = \tau/n$, such that $e^{-\imath \mathbf H\tau} \approx \left( e^{-\imath \mathbf H \tau/ n}\right)^n$. For small enough $\Delta \tau$, the following Suzuki-Trotter approximation of the second order applies~\cite{suzuki1993improved}:

\begin{align*}
    e^{-\imath(\mathbf H_1+ \mathbf H_2+\dots+ \mathbf H_k) \Delta \tau}& = \\e^{-\imath \mathbf H_1 \Delta \tau/2} &e^{-\imath \mathbf H_2 \Delta \tau/2} \cdots e^{-\imath \mathbf H_k \Delta \tau}  \cdots \\\cdots &e^{-\imath \mathbf H_2 \Delta \tau/2} e^{-\imath \mathbf H_1 \Delta \tau/2} + \mathcal O((\Delta\tau)^3)\text.
\end{align*}

For the FHM, we will first divide the Hamiltonian into three terms $\mathbf H = \mathbf H_U + \mathbf H_t + \mathbf H_{t'}$ corresponding to onsite interaction $\mathbf H_U = U\sum\limits_{j = 1}^{{L}} \mathbf n_{j, \uparrow} \mathbf n_{j, \downarrow}$ , nearest neighbour hopping $ \mathbf H_t = -t \sum\limits_{\sigma, j=1}^{{L}}\left[ \mathbf c_{j,\sigma}^\dagger \mathbf c_{j + 1, \sigma} + \text{H.c.} \right]$ and next-t-to-nearest hopping $ \mathbf H_{t'} = -t' \sum\limits_{\sigma, j=1}^{{L}}\left[ \mathbf c_{j,\sigma}^\dagger \mathbf c_{j + 2, \sigma} + \text{H.c.} \right]$ respectively. In this case, the evolution operator can be approximated: $e^{-i\Delta\tau \mathbf H} = e^{-i\Delta\tau \mathbf H_U} e^{-i\Delta\tau \mathbf H_t} e^{-i\Delta\tau \mathbf H_{t'}} \left(1+\mathcal O(\Delta\tau)\right)$. The Suzuki-Trotter approximation of the second order is then applied to the exponents of each term separately. 

Another approach to implementing dynamics in the quantum computing setting is based on the well-established variational quantum time evolution algorithm~\cite{Yuan2019}. This allows reconstructing the dynamics of a parameterized state encoded in a quantum circuit by solving an equation of motion of the circuit parameters. Advantages and disadvantages of this other approach will be discussed in appendix~\ref{Appendix:VariationalDynamics}.

\section{Results and Discussion} \label{Results and Discussion}
In this section, we apply the algorithms introduced above to the study of a $6$-site Hubbard model with nearest and next-to-nearest hopping terms. 

\subsection{Ground state and its properties} \label{Results.GroundState}
We begin with the calculation of the ground state wavefunction and properties for the model introduced in Eq.~\eqref{eq:Hamiltonian} restricted to nearest and next-to-nearest hopping. We benchmark our method against exact diagonalization and for different values of next-to-nearest hopping coefficient $t'$ and onsite interaction energy $U$. 

\begin{figure}[h!]
    \centering
    \includegraphics[width=\linewidth]{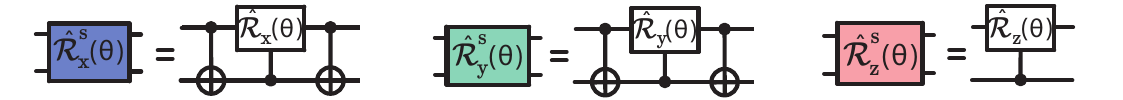}
    \caption{Circuit representation of symmetrized controlled rotations. The symmetrized rotations $\hat{\mathcal{R}}^{\text{s}}_{x,y,z} (\theta)$ are obtained by sandwiching the controlled $x, y, z$ rotations between two CNOT operations. $\hat{\mathcal{R}}^{\text{s}}_{x} (\theta)$ and $\hat{\mathcal{R}}^{\text{s}}_{y} (\theta)$ can be implemented in a more efficient manner using just two two-qubit gates (see Appendix~\ref{Appendix:Gates}).}
    \label{fig:Gates}
\end{figure}

To construct a VQE ansatz that respects the relevant symmetries of the system, we notice that the target Hamiltonian conserves the number of particles and the total spin of the system. For this reason, we use a set of symmetrized controlled rotations $\hat{\mathcal{R}}^{\text{s}}_x (\theta)$, $\hat{\mathcal{R}}^{\text{s}}_y (\theta)$ and $\hat{\mathcal{R}}^{\text{s}}_z (\theta)$ that commute with the total number operator $\mathbf N$ (but not with $\mathbf H$ globally) and thus preserve the particle number. The corresponding gates are shown in Fig.~\ref{fig:Gates}. Importantly, $\hat{\mathcal{R}}^{\text{s}}_z (\theta)$ preserves not only the total number of particles but also the local occupation on each of the qubits it is applied to. We will leverage this fact to introduce entanglement between the spin up and spin down sectors without changing the projection of the total spin along the $z$-axis of the system (i.e., perpendicular to the plane of the ring).

\begin{figure}[h!]
    \centering
    \includegraphics[width=1.0\linewidth]{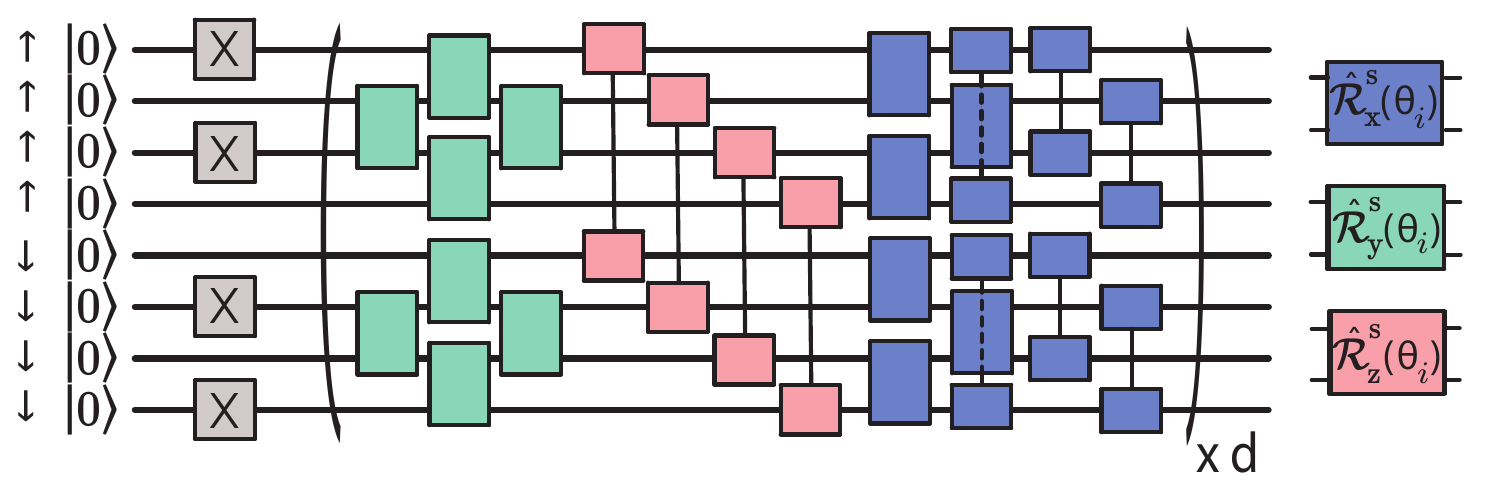}
    \caption{Structure of the HVA at half-filling for 4 lattice sites (8 qubits) and depth $d$. The number of electrons is set by the initial $X$ gates and is preserved along the circuit. $\hat{\mathcal{R}}^{\text{s}}_{x,y,z} (\theta)$ are the particle-preserving gates from Fig.~\ref{fig:Gates}.}
    \label{fig:Ansatz}
\end{figure}

For the ground state computation, we initialize the system state in the half-filled sector and further use gates that closely emulate the different interactions in the Hamiltonian. The resulting circuit structure is shown in Fig.~\ref{fig:Ansatz} and is an instance of Hamiltonian Variational Ans\"atze (HVA)~\cite{Wecker2014}, which are characterized by resiliency to barren plateaus and the ability to reproduce features of strongly-correlated systems with relatively few parameters~\cite{Wiersema2020, Jattana2022, Kattemolle2022, Bosse2021}, as demonstrated in~\cite{stanisic2022observing} for FHM restricted to nearest-neighbour interactions. We emphasize that the ansatz we use is inspired by, and extends upon, the one proposed in that work.

In more detail, we start from the reference state $|0\rangle^{\otimes n}$ and we first apply $X$ gates every second qubit to prepare an initial state with the correct number of particles with zero projection of the total spin along the z-axis (see Fig.~\ref{fig:Ansatz}).  

By placing spin up particles at odd sites and spin down at even sites, we prepare an antiferromagnetic initial state, which is known to be a good solution in the strongly interacting regime. We then apply $d$ layers of $\hat{\mathcal R}_y^{\text s}$ gates, followed by $\hat{\mathcal R}_z^{\text s}$, and finally symmetrized $x$-rotations $\hat{\mathcal R}_x^{\text s}$, each described by an indpendent rotation parameter $\theta_i$. Here, $\hat{\mathcal R}_y^{\text s}$ and $\hat{\mathcal R}_x^{\text s}$ represent Givens rotations (see Appendix~\ref{Appendix:Gates} for their matrix representation and for their decomposition in terms of CNOT operations). 

The optimization procedure goes as follows: (i) The circuit parameters are initialized to prepare the noninteracting ($U=0$) solution, i.e., a Slater determinant expressed in terms of Givens rotations. As previously noted in Ref.~\cite{zhang2018}, we observe that optimal convergence is achieved by using the fewest number of layers necessary to attain the desired accuracy. Conversely, overparameterization can significantly hinder convergence; (ii) a first VQE iteration is performed where all parameters  corresponding to nearest neighbour hopping, next-to-nearest neighbour hopping and onsite interactions in each layer are assigned common values in groups; (iii) a second VQE iteration is performed, starting from the parameters obtained in (ii),  this time allowing all parameters except those of the onsite interactions in each layer to vary independently. 

\begin{figure}[h!]
    \centering
    \includegraphics[width=\linewidth]{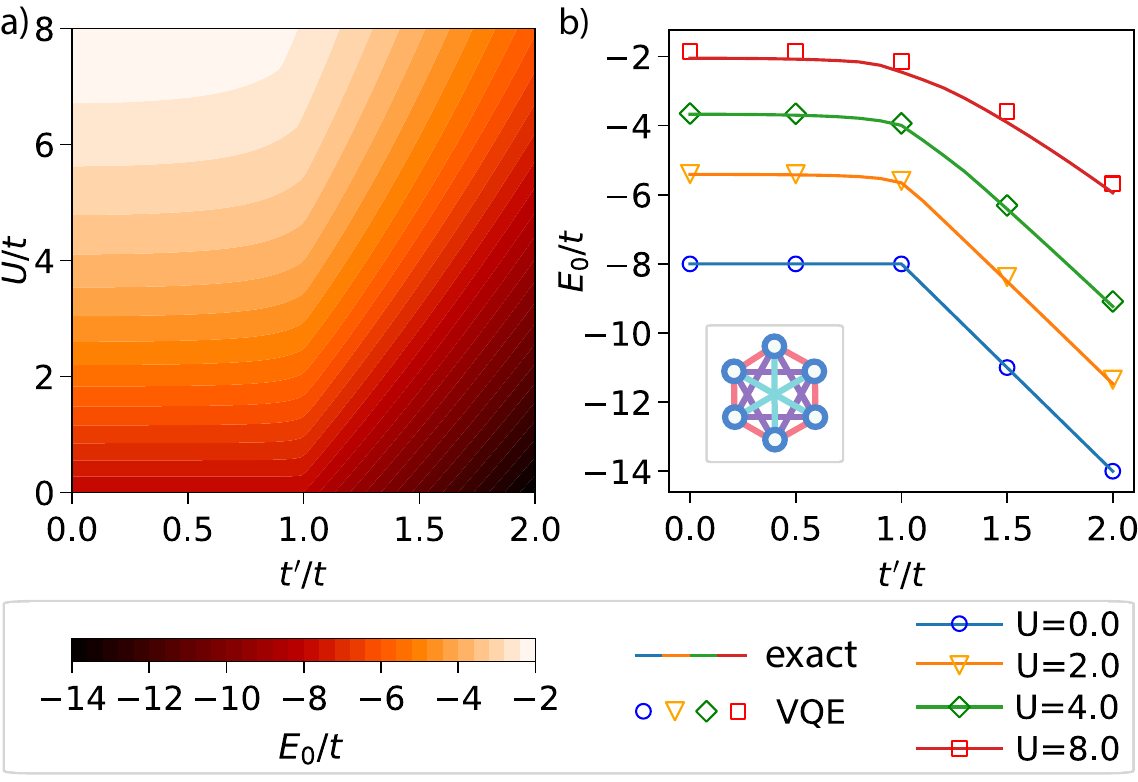}
    \caption{
        (a) Contour plot of the ground state energy $E_0$ as a function of the next-to-nearest neighbour hopping term $t'$ and the onsite interaction energy $U$, with $t$ representing the nearest neighbour hopping term. The plot is generated using exact diagonalizations performed with the NumPy package. The calculation is done for $L=6$ Hubbard sites in the half-filling regime (number of particles $N=L$).
        (b) Comparison of groundstate energies $E_0/t$ vs $t'/t$ as computed by exact diagonalization (continuous lines) and VQE (empty symbols) for different values of $U/t$ (see legend). VQE data are obtained with $d=4$ circuit repetitions in the absence of noise, using the SLSQP optimizer and a maximum of 1000 iterations.}
    \label{fig:GroundEnergies} 
\end{figure}

Figure~\ref{fig:GroundEnergies}a shows the contour plot of the ground state energy $E_0$  in units of nearest neighbours hopping $t$, computed with exact diagonalization as a function of next-to-nearest hopping coefficient $t'$ and on-site interaction energy $U$. The figure shows that the values of the ground state energies for any $t'/t<1$ are almost independent of the next-to-nearest hopping coefficient $t'$, while the structure of the ground state changes for $t'/t > 1$, where the increase in the next-to-nearest hopping decreases the energy almost linearly.

Figure~\ref{fig:GroundEnergies}b shows a few representative cuts through the surface in Fig.~\ref{fig:GroundEnergies}a, providing a better quantitative description of the underlying transition. Full lines represent matrix diagonalization results, while dots correspond to the values obtained from VQE. Round, diamond, triangular and square symbols (in blue, yellow, green and red colours) correspond to the the values of on-site interactions: $U/t=0.0, 2.0, 4.0, 6.0$, respectively. We observe that the VQE results are in good agreement with the reference values. In addition, for all values of on-site interaction $U$ we notice a sharp change in slope at roughly $t'/t=1$, which is an indication that, despite the limited number of sites in our calculations, it is already possible to detect the signature of phase transitions. We will come back to this point in Sec.~\ref{Results.Gaps}, where we discuss the thermodynamic limit of the model. 

\subsection{Charge and spin gaps} \label{Results.Gaps}
In this section, we investigate the appearance of charge and spin gaps, which play a fundamental role in determining the electronic properties and the overall behavior of the system. This analysis is crucial for understanding transport phenomena and the magnetic properties of the material under various conditions. 

We define the charge gap, $\Delta_{\text C} $, as the energy difference between the sum of the ground state energies of the systems with an extra particle and an extra hole, and the energy at half filling, $N_{\uparrow}=N_{\downarrow}=L/2$,

\begin{align*}
    \Delta_{\text C} = &E\left(N_\uparrow={\frac L2 + 1}, N_\downarrow=\frac L2\right) + \\ &E\left(N_\uparrow=\frac L2 - 1, N_\downarrow=\frac L2\right) - \\ &2E\left(N_\uparrow=\frac L2, N_\downarrow=\frac L2\right)\text.
\end{align*}

In a similar spirit, the spin gap, $\Delta_S$, is defined as the difference between the ground state energy of the 2-spin-flipped system and the one of the zero total spin state, 

\begin{align*}
    \Delta_{\text S} = &E\left(N_\uparrow=\frac L2 + 2, N_\downarrow=\frac L2 - 2\right) -\\ &E\left(N_\uparrow=\frac L2, N_\downarrow=\frac L2\right)\text.
\end{align*}

We measure these observables on the optimized VQE states and compare with the results from exact diagonalizations. To obtain the latter, we control the number of particles with spin up and spin down by adding to the Hamiltonian an additional term $-\lambda(\hat N_{\uparrow} - N_\uparrow)^2-\lambda(\hat N_{\downarrow} - N_\downarrow)^2 $, where $\hat N_{\uparrow(\downarrow)}$ is the operator of number of particles with spin up (down), $N_{\uparrow (\downarrow)}$ is the desired number of particles with spin up (down), and $\lambda$ is the Lagrange multiplier that we choose to be $10$ in our computations. 

In the VQE, instead, we leverage the fact that the proposed ansatz is both particle- and spin-preserving. Therefore we control the number of particles and total spin of the system wavefunction by initialising the ansatz in the desired symmetry sector. 

\begin{figure}[h!]
    \centering
    \includegraphics[width=\linewidth]{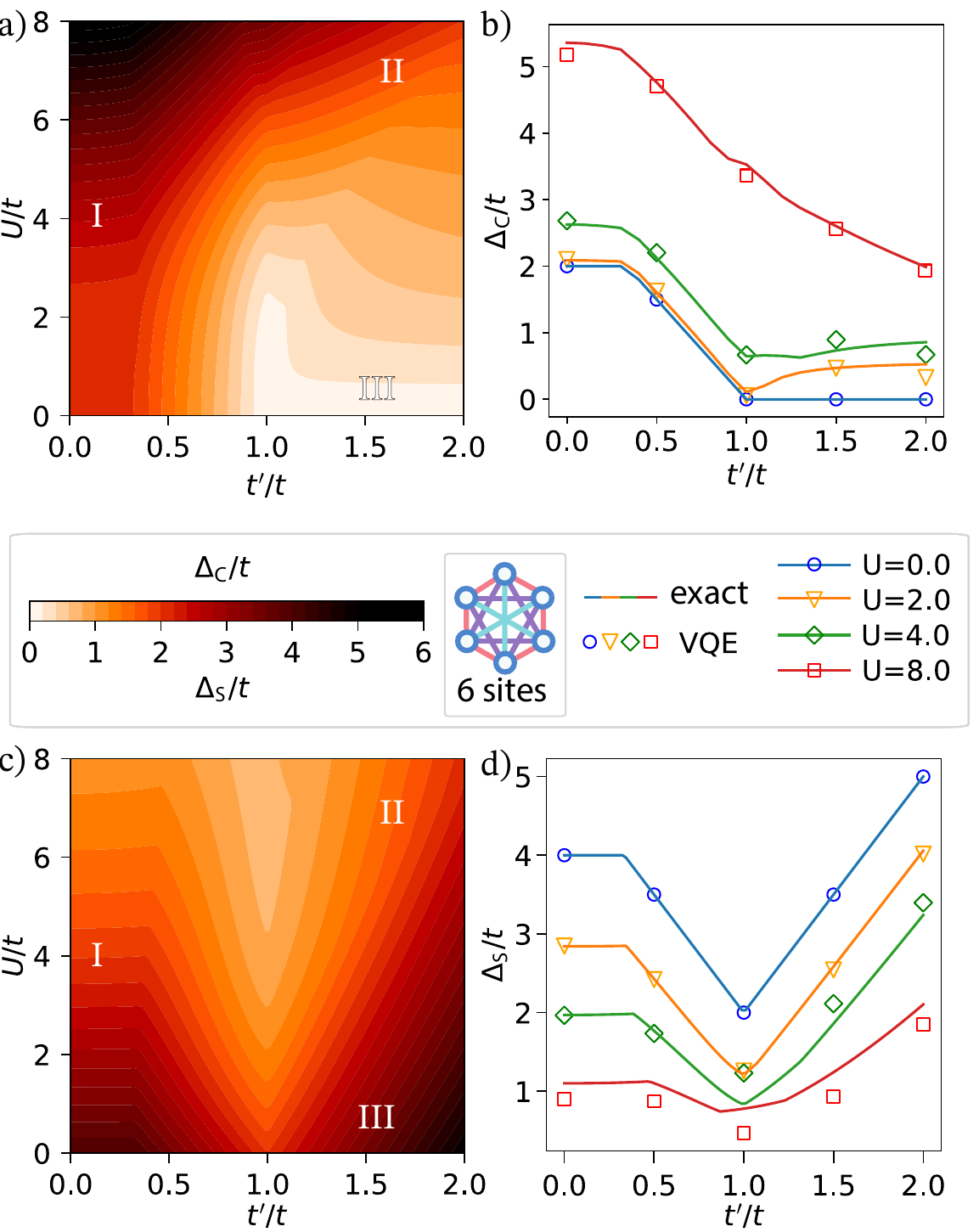}
    \caption{(a, c) Contour plots of the dimensionless charge gaps $\Delta_\text{C}$ (a) and the spin gaps $\Delta_\text{S}$ (c) as a function of the next-to-nearest hopping coefficient $t’$ and the onsite interaction energy $U$, with $t$ the nearest neighbour hopping coefficient. The results are obtained by exact diagonalization of the Hubbard Hamiltonian with $L=6$ sites in the half-filling regime (number of particles $N=L$) using the \texttt{numpy} package. 
    (b, d) Comparison of charge gaps $\Delta_\text{C}$ (b) and spin gaps $\Delta_\text{S}$ (d) vs $t'/t$ as computed by exact diagonalization (continuous lines) and VQE (empty symbols) for different values of $U/t$ (see legend). VQE data are obtained with $d = 6$ circuit repetitions  using the SLSQP optimizer and a maximum of 1000 iterations.}
    \label{fig:Gaps}
\end{figure}

Figure~\ref{fig:Gaps} summarizes the results obtained for the energy and spin gaps for $L=6$ sites. Panels a) and c) show contour plots of the charge gaps and spin gaps values, respectively, in the units of nearest-neighbour hopping $t$ as a function of next-to-nearest hopping coefficient $t'$ and on-site interaction energy $U$. As before, in panels b) and d) we also report a few slices evaluated at 4 different values of the interaction coefficient $U$. Full lines represent exact diagonalization results, while dots correspond to the VQE values. Round, diamond, triangular and square dots and blue, yellow, green and red colours correspond to the values of on-site interaction $U/t=0.0, 2.0, 4.0, 6.0$, respectively.
The labels I, II, III in Figs.~\ref{fig:Gaps}a,~\ref{fig:Gaps}c refer to the main phases expected in the thermodynamic limit~\cite{tocchio2010interaction}: spin gapless, spin and charge gapped, and charge gapless phase, respectively.
Region III in panels a) and c) corresponds to the metallic phase in the thermodynamic limit. Despite the relatively small number of sites used in our simulations, we can observe already for $L=6$ a charge gap approaching $0$. Similar observations apply to the regions I and II, which correspond to a charge-gapped phase in the thermodynamic limit and are well captured by our results for $L=6$.

Finally, regions II and III are predicted to describe a spin-gapped phase in infinitely large systems, and this is also the case for our simulations (Fig.~\ref{fig:Gaps}c). However, we also observe that, while a clear deep of the dip in the value of the gap is clearly visible at $t'/t=1$ as an indication of the transition, the spin-gap in region I does not approach zero but rather flattens to a sizable constant value at $t'/t=0$. This is a clear indication of finite-size effects, as it is the case for the shift of the transition point from the expected value at $t’/t=0.5$~\cite{tocchio2010interaction} to the observed one at $t’/t=1$.  A short explanation for this expected discrepancy is given in Appendix~\ref{Appendix:Transition}.

\subsection{Excited states and spectral function} \label{Results.SpectralFunction}
Moving on from the analysis of the ground state, we now consider the calculations of excited states' properties. To this end, we target the evaluation of the spectral function $A(\omega, k)$, an important physical quantity directly accessible in experiments. 
This can be obtained from the imaginary part of the single-particle retarded Green's function as $A(\omega, k) = -\text{Im }G^{\text{r}}_{k=0} (\omega) / \pi $. In turn, the diagonal terms of Green's function can be calculated using the Lehmann representation:
\begin{equation}
\begin{split} 
    \Tilde{G}^{\text{r}}_{\alpha\beta} (\omega) = \sum_n &\left( \frac{\bra{\psi_0}\mathbf c_\alpha\ket{E^{L+1}_n} \bra{E^{L+1}_n}\mathbf c_\beta^\dagger\ket{\psi_0}}{\omega + i\eta + E^{L}_0 - E^{L+1}_n} \right. \\ &~~+ \left. \frac{\bra{\psi_0}\mathbf c_\beta^\dagger\ket{E^{L-1}_n} \bra{E^{L-1}_n}\mathbf c_\alpha\ket{\psi_0}}{\omega + i\eta + E^{L-1}_n - E^{L}_0} \right)\text, 
\end{split}
\end{equation}
where $E_n^X$ is $n$-th excited state among the states with $X$ particles and $\eta$ is a damping factor, which in an experiment would correspond to energy loss in the system. The first term in the brackets corresponds to excited states with an extra particle, while the second to excited states with extra holes. In the following, we will call particle-like peaks the ones that contain a significant contribution from the first term, and hole-like peaks those that are more impacted by the second. 

To compute excited energies and wavefunctions we use the qEOM approach described in Sec.~\ref{Methods.qEOM}. As a pool of charging and de-charging excitation operators we use single excitations ($\mathbf c_{i\uparrow}^\dagger$ for $E^{L+1}_n$ and $\mathbf c_{i\uparrow}$ for $E^{L-1}_n$) and double excitations ($\mathbf c_{i\uparrow}^\dagger \mathbf c_{i'\uparrow}^\dagger \mathbf c_{j\uparrow}$ for $E^{L+1}_n$ and $\mathbf c_{i\uparrow}^\dagger \mathbf c_{j\uparrow} \mathbf c_{j'\uparrow}$ for $E^{L-1}_n$) operators. While this choice of operators cannot produce all possible excited states, it allows for the computation of several low-lying excited states.

\begin{figure}[h!]
    \centering
    \includegraphics[width=\linewidth]{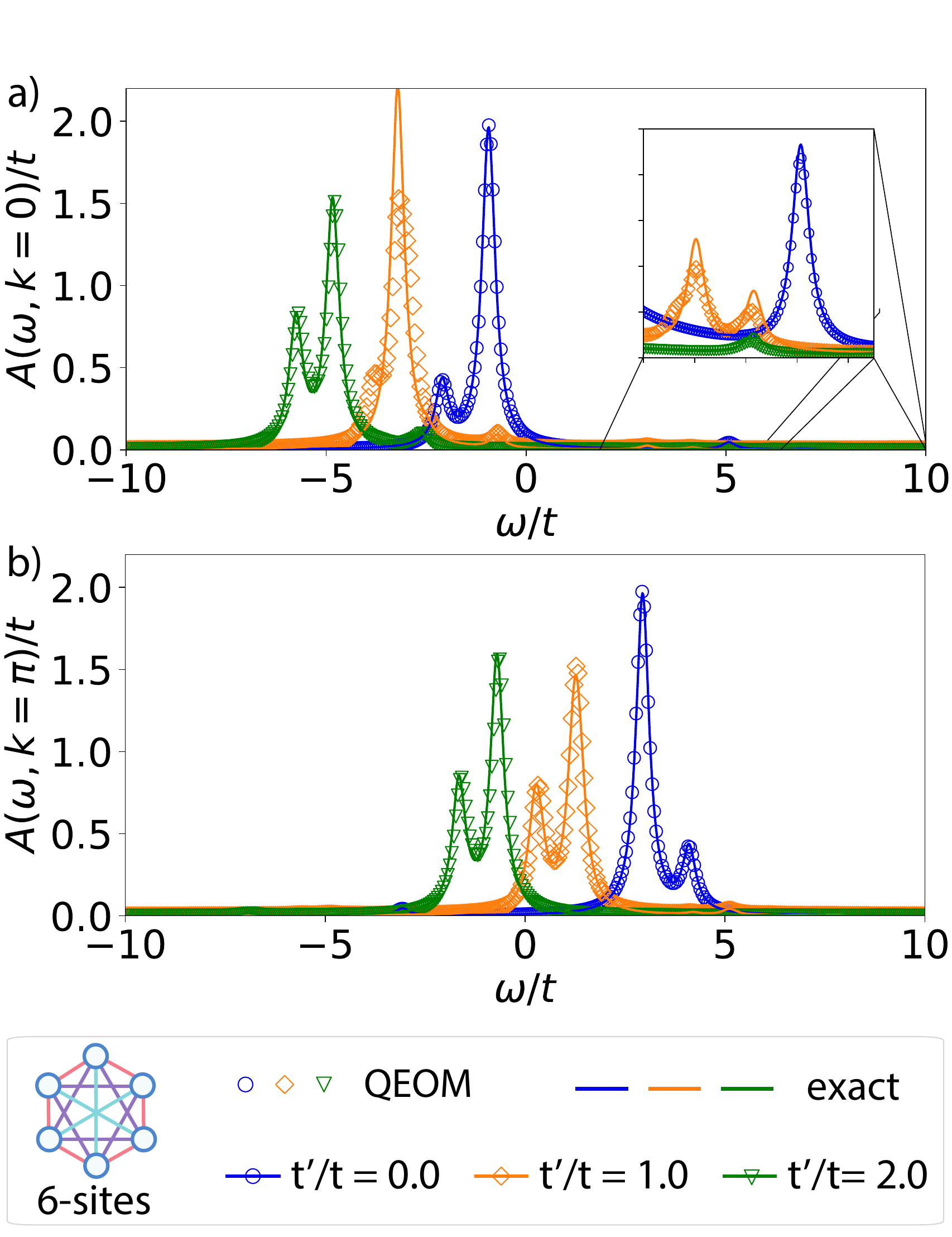}
    \caption{Spectral function $A(\omega, k)$ as a function of frequency $\omega$ for two values of $k$. All quantities are measured in units of the nearest neighbour hopping term, $t$.
    The results are for the Hubbard model with $L=6$ sites in the half-filling regime (number of particles $N=L$), an onsite interaction energy $U=2t$, and different values of next-to-nearest hopping energy $t'$ (see legend). 
    The calculations are performed using the Lehmann representation in $k$ space for $k=0$ (a) and $k=\pi$ (b) with damping factor $\eta = 0.2$. Lines correspond to exact diagonalization, dots  to the qEOM results obtained after ground state optimization using VQE with $d = 4$ repetitions and a convergence threshold of $5\cdot 10^{-3}$.}
    \label{fig:SpectralFunction}
\end{figure}

Fig.~\ref{fig:SpectralFunction} summarizes the results of the spectral function calculations in the Mott insulator phase for $L=6$ Hubbard sites, and the value of on-site interaction $U=2.0t$. The different symbols and colours represent different values of next-to-nearest hopping terms. Green, orange and blue (and equivalently triangular, diamond and circle symbols) correspond to $t' = 0.0 t, 1.0 t, 2.0 t$ values, respectively. The spectral function for momentum $k=0$ is shown in panel a) and for momentum $k=\pi$ in panel b); in both cases, we have chosen a damping factor $\eta=0.2$. For the calculation of matrix elements required in Eq.~\eqref{eq:GeneralizedEigenvalue}, we consider VQE states prepared with a fidelity $F \geq 0.995$. To improve the stability of the qEOM approach, the generalized eigenvalue problem is solved after applying a series of filtering functions: first, we only consider states with $\lVert X_{\mu}\rVert > 10^{-2}$, $\lVert \hat O^\dagger_n \ket{0} \rVert > 10^{-3}$, and a total number of particles within $10^{-3}$ from the half-filling condition; second, for the states within degenerate multiplets, we perform an extra orthogonalisation procedure. 
The spectrum is made of a double-peak structure that shifts to lower frequencies with increasing $t'$ (Fig.~\ref{fig:SpectralFunction}a). The distance between the sub-peaks is roughly proportional to the onsite interaction $U$ while the centre of the structure is roughly at the position of the corresponding ground energy of the non-interacting model (Appendix~\ref{Appendix:Transition}). Peaks at higher energies (shown in the inset) correspond particle-like ones at energies around $2U$. We observe that when moving from $k=0$ (panel a) to $k=\pi$ (panel b) the overall structure of the peaks remains unchanged, but their position shifts by roughly $2U$. 

\subsection{Dynamics} \label{Results.Dynamics}
Of particular interest for the characterization of the physical properties of the FHM with long-range hopping terms is the investigation of dynamic properties such as dynamic correlations and dynamic structure factors. In particular, in this work we are focusing on the calculation of spin-spin correlation functions~\cite{chiesa2019quantum}: 
\begin{align*}
    C^{zz}_{ij}(t) =  \langle \sigma^z_i(t) \sigma^z_j(0) \rangle\text,
\end{align*}
which can be further used to determine the corresponding structure factor:
\begin{align*}
    S_{zz}(q, \omega) = 
    \frac{1}{L}& \sum_{i, j} \int_{0}^T\mathrm dt\; e^{-iq(x_i-x_j)} e^{i\omega t} \langle \sigma^z_i(t) \sigma^z_j(0) \rangle \text,
\end{align*}
where $\sigma_i$ is the spin operator on site $i$.

To perform the dynamics, we first compute the ground state by applying the VQE as in Sec.~\ref{Results.GroundState} and then propagate the state by implementing the second-order Trotter approximation of the time evolution operator as a circuit~\cite{Miessen2023} as described in Sec.~\ref{Methods.Trotterization}. (The quantum simulation circuits for the $L=6$ case are provided in Appendix~\ref{Appendix:Trotter}.)

\begin{figure}[h!]
    \centering
    \includegraphics[width=0.75\linewidth]{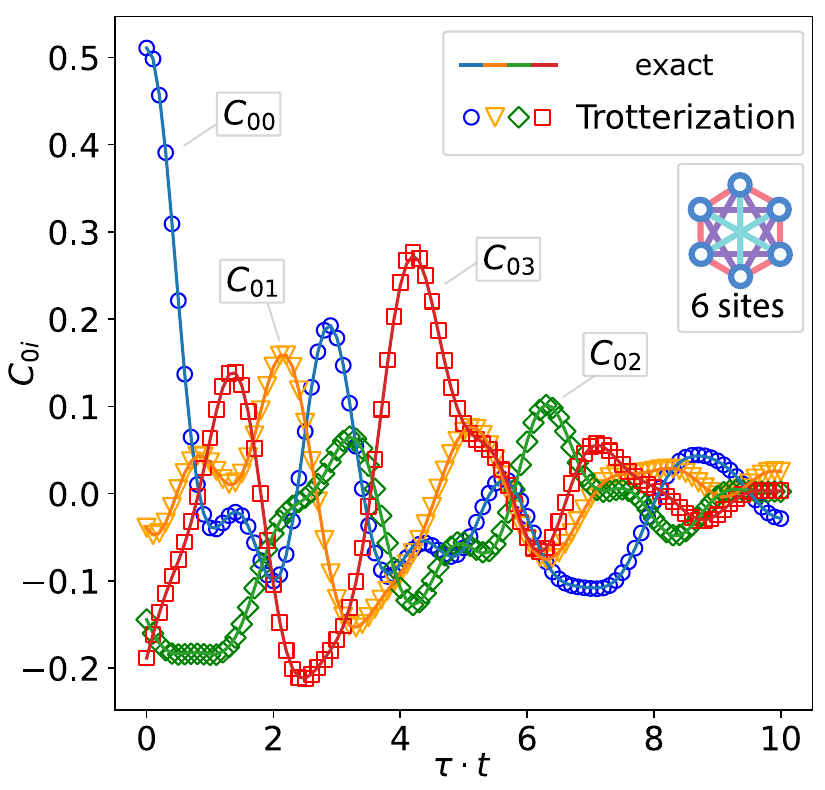}
    \caption{Dynamical spin-spin correlations $C_{0j}$ as a function of time $\tau$ for $L=6$ Hubbard sites in the half-filling regime (number of particles $N=L$), with onsite interaction energy $U/t=0.5$, and next-to-nearest hopping energy $t'/t=0.5$. Full lines correspond to the statevector results performed with \texttt{numpy}. The dots are the results obtained with the second-order Trotter circuit where a new Trotter layer was added every $\Delta\tau \cdot t = 0.1$. The system was initialized in the ground state using a VQE with a convergence threshold of $5\cdot 10^{-3}$ and a circuit ansatz composed of $d=4$ repetitions.}
    \label{fig:TrotterisationStepByStep}
\end{figure}

\begin{figure}[t!]
    \centering
    \includegraphics[width=\linewidth]{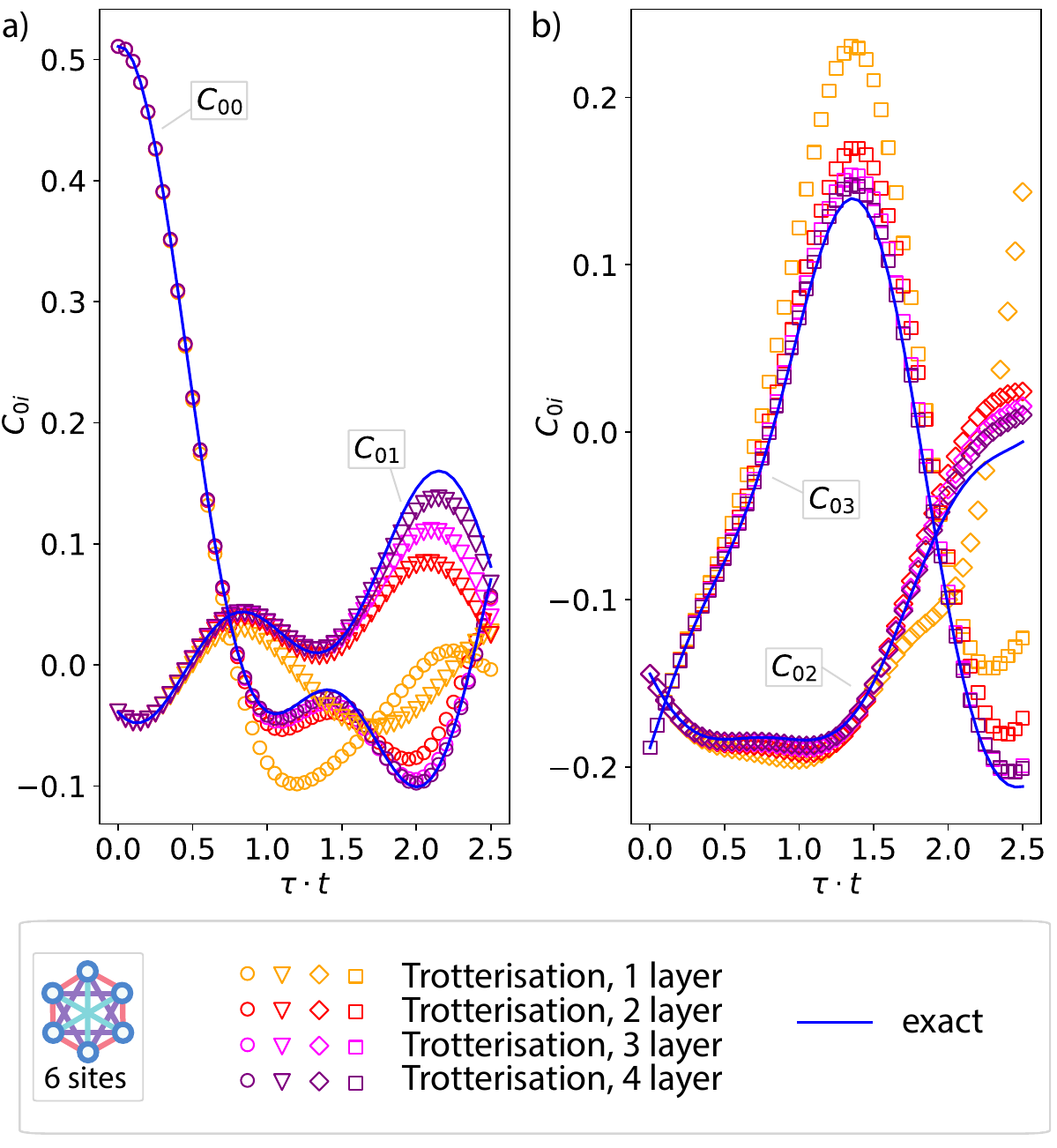}
    \caption{Dynamical spin-spin correlations $C_{0j}$ as a function of $\tau$ for the $L=6$ Hubbard model in the half-filling regime (number of particles $N=L$), with onsite interaction energy $U/t=0.5$, and next-to-nearest hopping energy $t'/t=0.5$. (a) Equal site ($C_{00}$) and nearest site ($C_{01}$) correlations, (b) next-to-nearest site ($C_{02}$) and opposite site ($C_{03}$) correllations. Full lines represent results from  statevector calculations performed with \texttt{numpy}, while dots correspond to values obtained with the second-order Trotter formula with 1, 2, 3, 4 repetitions of the corresponding circuit (see legend). 
    The system was initialized in the ground state using a VQE with a convergence threshold of $5\cdot 10^{-3}$ and a circuit ansatz composed of $d=4$ repetitions.}
    \label{fig:Trotterisation}
\end{figure}

Figs.~\ref{fig:TrotterisationStepByStep} and~\ref{fig:Trotterisation} show the dynamic spin-spin correlations $C_{ij}$ as a function of the delay time $\tau$ in units of $1/t$. In the first scenario (Fig.~\ref{fig:TrotterisationStepByStep}), we continuously add a new Trotter layer as we progress along the dynamics, each layer implementing a Trotter step of size $\delta \tau$. The different symbols (and colours) correspond to correlation functions between different sites 0 and $i$, with $i\in\{0,1,2,3\}$ (see labels of the curves).
As expected, after choosing an adequate time step ($\Delta t$) that guarantees the stability over the first time step, the dynamics is able to reproduce the exact profile computed by matrix diagonalization (continuous lines).
The main drawback of this approach is the linear increase in the circuit depth as a function of the total simulation times, which drastically limits the applicability of this approach on noisy quantum hardware. To overcome this limitation, we also tested the possibility of implementing a fixed number of Trotter layers while allowing $\Delta \tau$ to stretch between the initial and final time of the simulation. In this test we performed simulations with 1 to 4 layers, each represented by a different colour in Fig.~\ref{fig:Trotterisation}. 
As before, different symbols correspond to correlation functions between different sites as indicated by the labels on the curves. We observe that while adding a Trotter layer at each step we can reach a good accuracy over a total time of $\tau=10 \, t^{-1}$ and beyond, with a maximum of 4 Trotter step we can only reproduce the first a total time of $2 \, t^{-1}$ before the simulation start deviating significantly for the exact reference curve (Fig.~\ref{fig:Trotterisation}). It is nevertheless remarkable that, even with such rather shallow circuits made of only 4 Trotter layers, it is already possible to capture most of the high frequency oscillations of the correlation functions. However, we do not expect this behaviour to carry over in an effective way to larger system sizes.

It is worth mentioning that other time-propagation algorithms can be considered for the calculations of these  time-dependent observables, such as the quantum variational time-propagation approach~\cite{Miessen2023}. We provide a quick overview of the results obtained with this approach using two possible ans\"atze (the VQE ansatz and a single layer Trotter circuit) in Appendix~\ref{Appendix:VariationalDynamics}. A careful analysis of the application of the the variational time-propagation approach to our systems is beyond the scope of this work.

\subsection{Scaling} \label{sec:Scaling}
Here, we present the scaling of the number of gates required to construct the ansatze for both statics and dynamics, for growing target system size. We focus on the number of two-qubit gates, as they typically exhibit lower accuracy in experimental realizations, making their quantity the limiting factor in real-device calculations.

Table~\ref{tab:Scaling} reports on the scaling of the ground state VQE ansatz (top) and of the dynamics using the Trotter formula (bottom) as a function of the number of Hubbard sites $L$ and circuit repetitions $d$. For the static circuits, the different columns show the number of gates required to implement the $\hat{\mathcal R}^{\text{s}}_y$, $\hat{\mathcal R}^{\text{s}}_z$ and $\hat{\mathcal R}^{\text{s}}_x$ components of the static ansatz,  respectively. For the $\hat{\mathcal R}^{\text{s}}_x$ gates, we report separately the the number of CNOTs for the implementation of the nearest (N) and next-to-nearest (NN) terms. The lower part of the table summarizes column-wise the number of CNOT gates used to implement a single Trotter step ansatz decomposed into onsite interaction $U$, nearest hopping $t$ and next-to-nearest hopping $t'$ terms, respectively. The final mapping of the evolution operators into the corresponding quantum circuit is accomplished using the available Qiskit transpilation~\cite{Qiskit} feature without any further optimization. No additional overhead for the calculation of the correlation functions through, for instance, the Hadamard test~\cite{ekert2002direct, huang2021near, wu2021quantum, bravo2023variational, li2024efficient} is considered.

\begin{table}[t!]
    \centering
    \begin{tabular}{|c||c|c|c|c|} 
        \hline 
        Statics & $R^S_y$ &  $R^S_z$ & $R^S_x$ (N+NN) & Swap\\ \hline
        \#gates& $(3L-4)d$ & $Ld$ & $2Ld +2Ld$ & $(18L-32)d$\\ \hline
        \#(CNOTs) & $2(3L-4)d$ & $2Ld$ & $2(2Ld +2Ld)$ & $3(18L-32)d$\\ \hline
        $L=20$ & $112d$ & $40d$ & $160d$ & $984d$\\ \hline\hline
        Dynamics  & $U$ & $t$ & $t’$ & Swap\\ \hline
        \#(CNOTs) & $2L$ & $28 L - 36$ & $56 L  - 120$ & $\sim3(6L)$\\ \hline
        $L=20$ & $40$ & $524$ & $1000$ & $348$\\ \hline
    \end{tabular}
    \caption{
    Upper panel (Statics): Scaling of the number of symmetrized 2-qubit operations and swap operations as a function of the number of lattice sites $L$ and number circuit repetitions $d$ (for the VQE ansatz). `\# (gates)' stands for the count of gate operations while `\# (CNOTs)' corresponds to the total number of CNOTs required for the circuit implementation. 
    Lower panel (Dynamics): same count for the implementation of the $U$, $t$, and $t'$ terms of a single (second order) Trotter step. 
    In both cases, we also provide numerical values for the CNOT count for $L=20$. The number of SWAPs refers to the mapping reported in Fig.~\ref{fig:Mapping}a). 
    }
    \label{tab:Scaling}
\end{table}

\begin{figure}[h!]
    \centering
    \includegraphics[width=\linewidth]{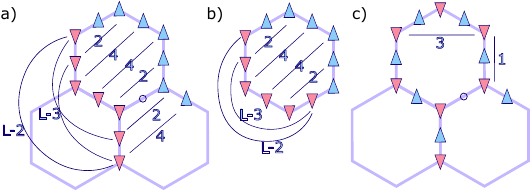}
    \caption{Possible mappings of a linear spin chain into a heavy-hex lattice. Filled up (down) triangles represent the location of spins up (down). Empty circles indicate the positions of extra device qubits, which are not part of the spin chain. The values on the black lines correspond to the number of swaps needed to bring the connected qubits on neighbour positions. (a, c) give mappings for $L=7$ sites, (b) for $L=6$ sites.}
    \label{fig:Mapping}
\end{figure}

To perform the proposed procedures on a real device, one must consider its architecture, and particularly its native qubit-qubit connectivity. Here, we examine the coupling map of IBM Quantum devices, which feature a heavy-hexagonal qubit lattice topology, namely a hexagonal lattice enhanced with an additional qubit on each edge. We compare several mappings and provide a straightforward computation of the upper limit on the number of swap gates required for each case. For a deeper discussion on optimizing the mapping, see Appendix~\ref{Appendix:Mapping}.

Figure~\ref{fig:Mapping}a,~\ref{fig:Mapping}c shows possible mappings of the qubits corresponding to up (blue up-triangles) and down (rose down-triangles) spin orbitals to the heavy-hex architecture of IBM Quantum devices. For the VQE ansatz that includes nearest and next-to-nearest hopping terms, the upper bound for the number of swap gates needed to implement the ansatz grows as $\mathcal{O}\left((18L - 28)d\right)$ for mapping a), and $22\, Ld$ for mapping b). We see that the leading term of the scaling is $22 \, Ld$ in both cases, while the variant a) is better by a constant term. In the $L=6$ case, we find the mapping in Fig.~\ref{fig:Mapping}b to be optimal with a total of $38 \, d $ swap gates needed.

The implementation of the mapping in Fig.~\ref{fig:Mapping}a will in general require twice the number of swap operations for each $\hat{\mathcal{R}}^{\text{s}}_{z} (\theta) $ layer of gates compared to case a) (twice the amount is needed to first make qubits close to one another and then to move them back to their initial places); this gives an overhead of $\mathcal{O}(6L)$ swap gates.
The same additional cost applies to the implementation of the onsite interaction term of the Trotter ansatz. For the $\hat{\mathcal{R}}^{\text{s}}_{y} (\theta) $ layer in the VQE ansatz and the hopping terms of the Trotter circuit there are no additional swaps needed, as in this case as nearest sites with same spin are arranged sequentially.

\section*{Conclusions}
This work investigated the application of quantum computing algorithms to the study of the static and dynamic properties of the 1D Fermi-Hubbard model with next-to-nearest neighbour hopping terms.

In particular, we considered the variational quantum eigensolver (VQE) for the calculation of the ground state wavefunction and properties, the quantum equation of motion (qEOM)  algorithm for the evaluation of excited states' energies and properties, and a product formula-based approach for the calculation of spectral functions and dynamic correlations. Our main results can be organized into five points.

(\textit{i})  We assessed the quality of the selected algorithms and circuit implementations by comparing the results obtained from the simulation of the proposed circuits with exact numerical values obtained from the exact diagonalization of the system's Hamiltonian. In all cases, we observe that the proposed circuit representation of the ground state ansatz and the time propagation unitaries used for dynamics can reproduce the reference calculations with the desired accuracy. In addition, by increasing the number of circuit layers we could prove a smooth convergence towards the exact results, providing therefore a systematic way of improving the performance of the quantum approaches.

(\textit{ii}) We also investigated the impact of the fine-size effects on the characterization of quantities that in principle would require simulations in the thermodynamic limit; we notice that while the results may in some cases quantitatively differ from those obtained by means of larger classical calculations, the main signatures for the presence of phase boundaries are nevertheless detectable also in our small scale quantum simulations.

(\textit{iii}) Regarding the dynamics, we first determined the largest time step compatible with our implementation of a single, second-order Trotter step. We then computed spin-spin time-dependent correlation functions using relatively shallow circuits, i.e., with a depth compatible with state-of-the-art quantum hardware, observing excellent agreement with the exact curves.

(\textit{iv}) For all applications, we also discussed various possible circuit implementations and analyzed how the circuit depth scales with the number of target system sites.

(\textit{v}) We provided an estimate of the resources, specifically the gate count, needed to reach system sizes that are not easily accessible by non-approximate classical solvers, such as exact diagonalization. As an explicit example, we considered a system with 20 sites (40 qubits), demonstrating that the ground state and dynamic properties for the FHM with next-to-nearest neighbour interactions will require fewer than 2,000 CNOT operations on a device with all-to-all connectivity and about 5,000 CNOT operations when mapped to the heavy-hexagonal topology of current IBM quantum hardware.

This work represents an initial step in evaluating the performance of quantum computing calculations on current and future generations of quantum hardware. Future studies should address the impact of hardware noise on measured observables and investigate their convergence as a function of system size, extending beyond the one-dimensional models analyzed in this study.

\section*{Acknowledgements}
This research has received funding from the European Union’s Horizon 2020 research and innovation programme under the Marie Skłodowska-Curie grant agreement number 955479 (MOQS - Molecular Quantum Simulations), the Horizon Europe programme HORIZON-CL4-2021-DIGITAL-EMERGING-01-30 via the project 101070144 (EuRyQa) and from the French National Research Agency under the Investments of the Future Program projects ANR-21-ESRE-0032 (aQCess). F.T. and I.T. were also supported by the NCCR MARVEL, a National Centre of Competence in Research, funded by the Swiss National Science Foundation (grant number 205602). IBM, the IBM logo, and ibm.com are trademarks of International Business Machines Corp., registered in many jurisdictions worldwide. Other product and service names might be trademarks of IBM or other companies. The current list of IBM trademarks is available at \url{https://www.ibm.com/legal/copytrade}.

\bibliography{bibliography.bib}

\appendix
\section{Efficient implementation and matrix form of $\hat{\mathcal{R}}^{\text{s}}_{x,y,z} (\theta)$ gates} \label{Appendix:Gates}
The matrix forms of the particle preserving gates introduced in Sec.~\ref{Results.GroundState}:

\begin{align*}
    \hat{\mathcal{R}}^{\text{s}}_y (\theta) &=  \left( \begin{matrix}
        1 & 0 & 0 & 0\\
        0 & \cos\left(\frac{\theta}{2}\right) & -\sin\left(\frac{\theta}{2}\right) & 0\\
        0 & \sin\left(\frac{\theta}{2}\right) & \cos\left(\frac{\theta}{2}\right) & 0\\
        0 & 0 & 0 & 1 
    \end{matrix} \right)\text,\\
    \hat{\mathcal{R}}^{\text{s}}_x (\theta) &= \left( \begin{matrix}
        1 & 0 & 0 & 0\\
        0 & \cos\left(\frac{\theta}{2}\right) & -i\sin\left(\frac{\theta}{2}\right) & 0\\
        0 & -i\sin\left(\frac{\theta}{2}\right) & \cos\left(\frac{\theta}{2}\right) & 0\\
        0 & 0 & 0 & 1 
    \end{matrix} \right)\text,\\
    \hat{\mathcal{R}}^{\text{s}}_z (\theta) &= \left( \begin{matrix}
        1 & 0 & 0 & 0\\
        0 & 1 & 0 & 0\\
        0 & 0 & \exp\left(-i\frac{\theta}{2}\right) & 0\\
        0 & 0 & 0 & \exp\left(i\frac{\theta}{2}\right) 
    \end{matrix} \right)\text.
\end{align*}

These gates allow us to obtain the form of the general gate that conserves the number of particles, which matrix form can be written as follows

\begin{align*}
    \left( \begin{matrix}
    e^{-i\phi_a} & 0 & 0 & 0\\
    0 & r_b e^{-i\phi_b} & (1-r_b)e^{-i\phi_c} & 0\\
    0 & (1-r_b) e^{-i\phi_d} & r_b e^{-i\phi_f} & 0\\
    0 & 0 & 0 & e^{-i\phi_g} 
    \end{matrix} \right)\text.
\end{align*}

\begin{figure}[!h]
     \[ \text{a)} \hspace{0.25cm} 
         \Qcircuit @C=1em @R=1em {
        &\gate{H}  &\ctrl{1} &\gate{\hat{\mathcal R}_y(\frac{\theta}{2})} &\ctrl{1} &\gate{H}  & \qw \\
        &\qw  &\targ & \gate{\hat{\mathcal R}_y(\frac{\theta}{2}) }& \targ & \qw & \qw  \\}
    \]\\
    \[\text{b)} \hspace{0.25cm} 
        \Qcircuit @C=1em @R=1em {
        &\gate{H}  &\ctrl{1} &\gate{\hat{\mathcal R}_x(\frac{\theta}{2})} &\ctrl{1} &\gate{H} & \qw  \\
        &\qw  &\targ & \gate{\hat{\mathcal R}_x(\frac{\theta}{2}) }& \targ & \qw & \qw  \\}
    \]\\
    \[\text{c)} \hspace{0.25cm} 
        \Qcircuit @C=1em @R=1em {
        & \qw  &\ctrl{1} &\gate{\hat{\mathcal R}_z(\frac{\theta}{2})} &\ctrl{1} & \qw  & \qw \\
        &\gate{X}  &\targ & \gate{\hat{\mathcal R}_z(\frac{\theta}{2}) }& \targ & \qw  & \qw \\}
    \]
    \caption{The efficient expansion of (a) $\hat{\mathcal{R}}^{\text{s}}_{x} (\theta)$, (b) $\hat{\mathcal{R}}^{\text{s}}_{y} (\theta)$, (c) $\hat{\mathcal{R}}^{\text{s}}_{z} (\theta)$ into CNOTs and single-qubit rotations}
    \label{fig:givens}
\end{figure}
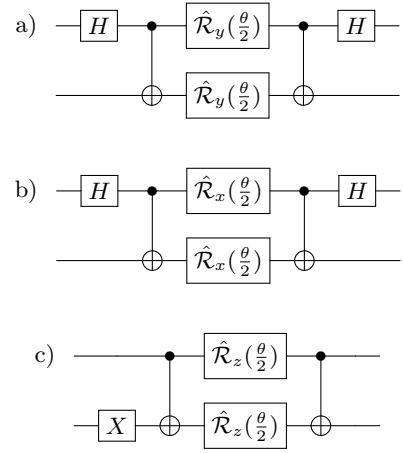

The efficient expansion of used gates into CNOTs and single-qubit rotations (as done in~\cite{jamet2022quantum} for Givens rotations) is shown in Fig.~\ref{fig:givens}

\section{Transition point} \label{Appendix:Transition}
We note that in the $L=6$ system the symmetry of the ground state changes at $t'=1.0$ while in the thermodynamic limit it happens at $t'=0.5$. Here we provide a brief explanation of why this happens.

\begin{figure}[H]
    \centering
    \includegraphics[width=\linewidth]{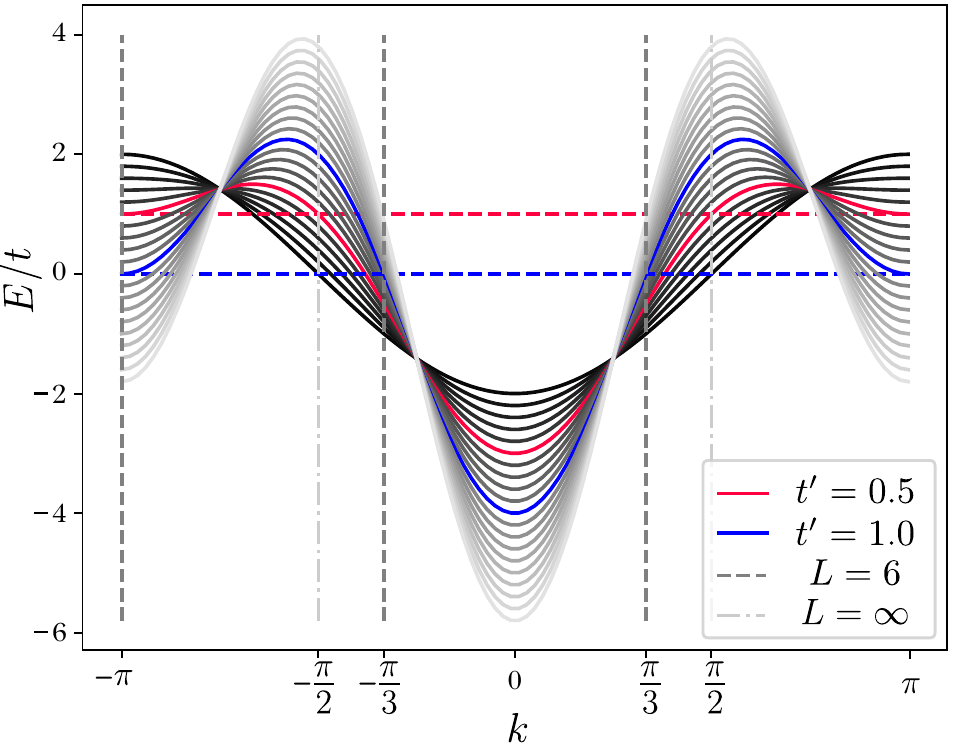}
    \caption{Energy levels $E$ in the units of nearest-neighbour hopping energy $t$ as a function of the wave-vector $k$. Vertical dashed lines mark the half-filling (dark grey for $L=\infty$, light grey for $L=6$). Horizontal dashed lines mark the instances where a particle at $k=\pi$ and at the edge of central half-filling zone have the same energy.}
    \label{fig:Energylevels}
\end{figure}

In Fig.~\ref{fig:Energylevels} we show the band diagram for the Fermi-Hubbard system without on-site interaction ($U=0$) for different values of the next-to-nearest hopping term $t'$ (the darkest line corresponds to $t'/t=0$, the lightest - to $t'/t=1.9$). The ground state qualitatively changes when with an increase in the next-to-nearest hopping term $t'$ some of the particles from being localized in the central part of the diagram start being localized at the edges of the diagram at $k=\pi$. For $L=6$ and $L\to\infty$ this occurs for different values of $t'$. In the diagram we highlighted the corresponding energy levels for when it happens in half-filling, in red for $L\to\infty$ and blue for $L=6$.

\section{Trotter circuits} \label{Appendix:Trotter}
\begin{figure}[H]
    \centering
    \includegraphics[width=\linewidth]{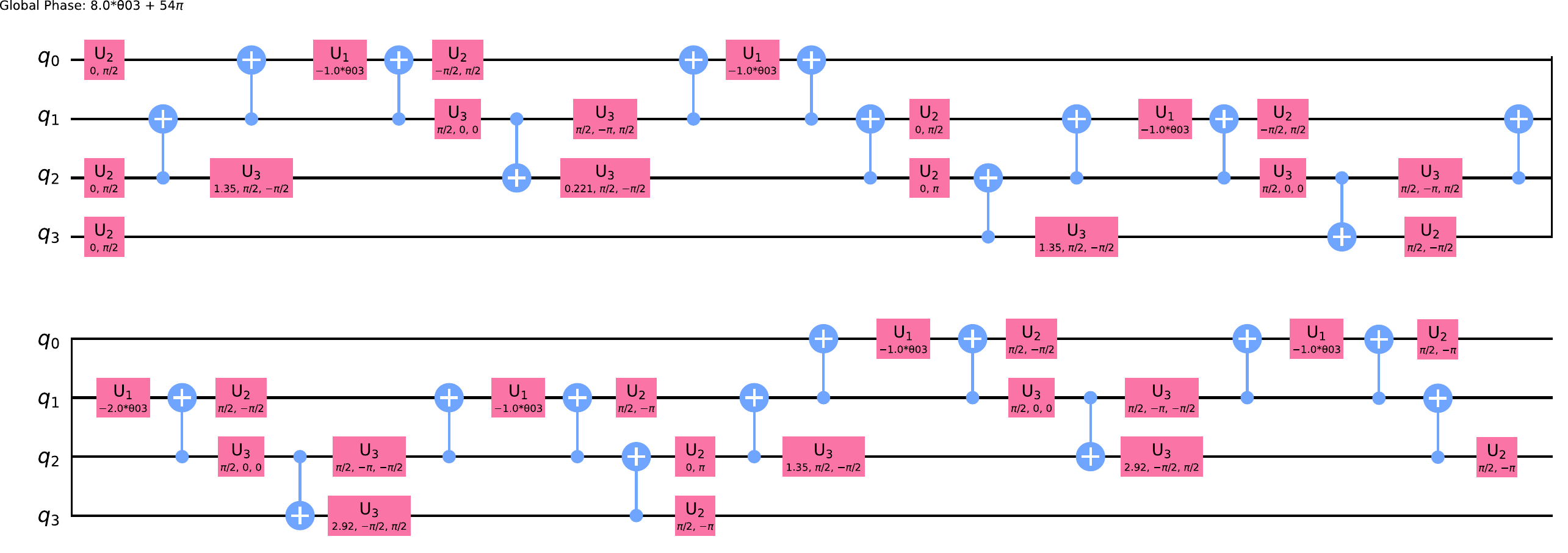}
    \caption{Next-to-nearest hopping part of Trotter circuit used in Sec.~\ref{Results.Dynamics} shown for the qubits corresponding to spins up only (the exact same structure applies to the qubits corresponding to spins down).}
    \label{fig:Trotter_tdash}
\end{figure}

In the following figures~\ref{fig:Trotter_U},~\ref{fig:Trotter_t}, and ~\ref{fig:Trotter_tdash}, we provide the explicit form of the Trotter circuits for the example with $L=4$ Hubbard sites. 

\begin{figure}[H]
    \centering
    \includegraphics[width=\linewidth]{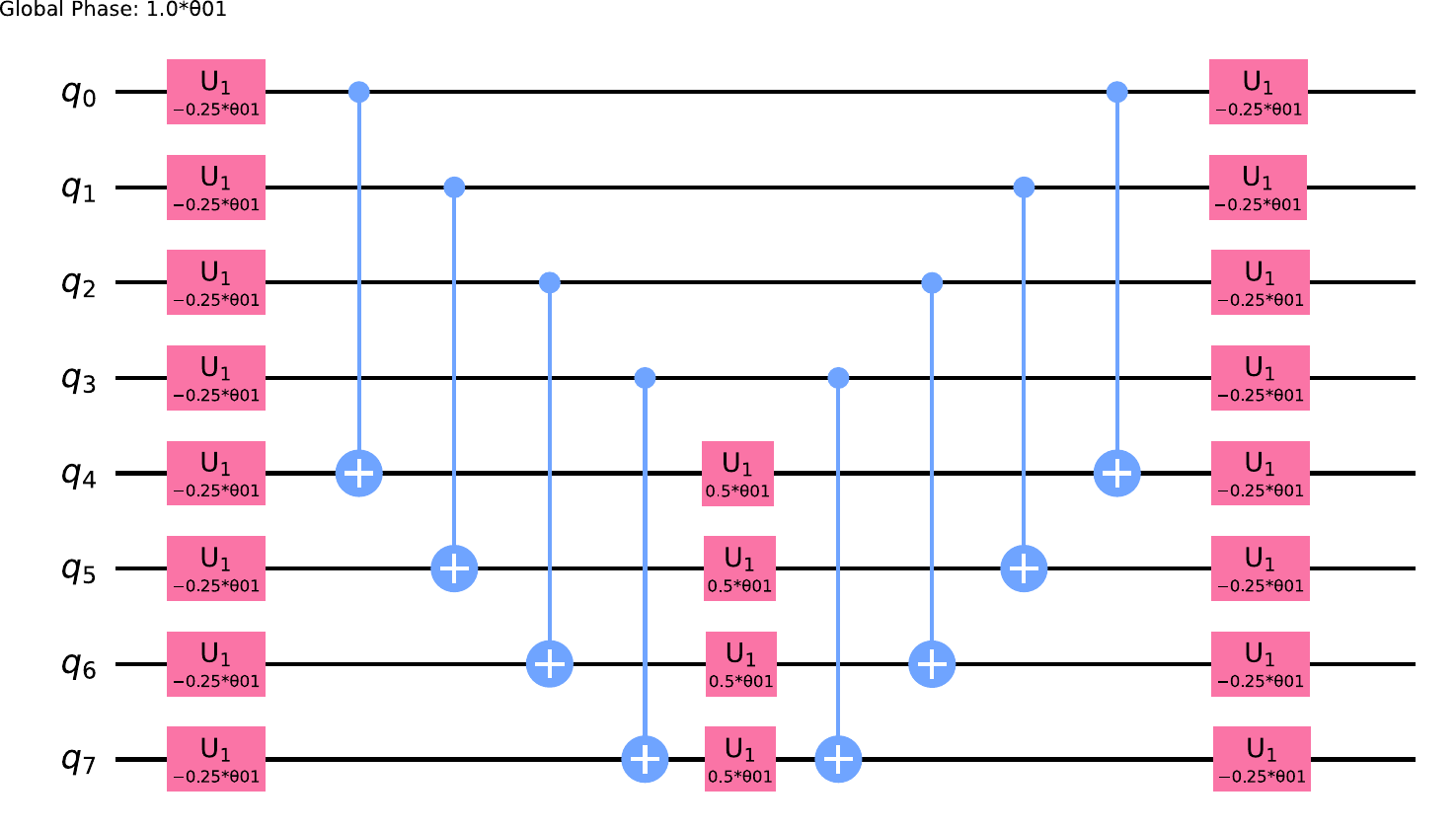}
    \caption{On-site interaction part of Trotter circuit used in Sec.~\ref{Results.Dynamics}.}
    \label{fig:Trotter_U}
\end{figure}

\begin{figure}[H]
    \centering
    \includegraphics[width=\linewidth]{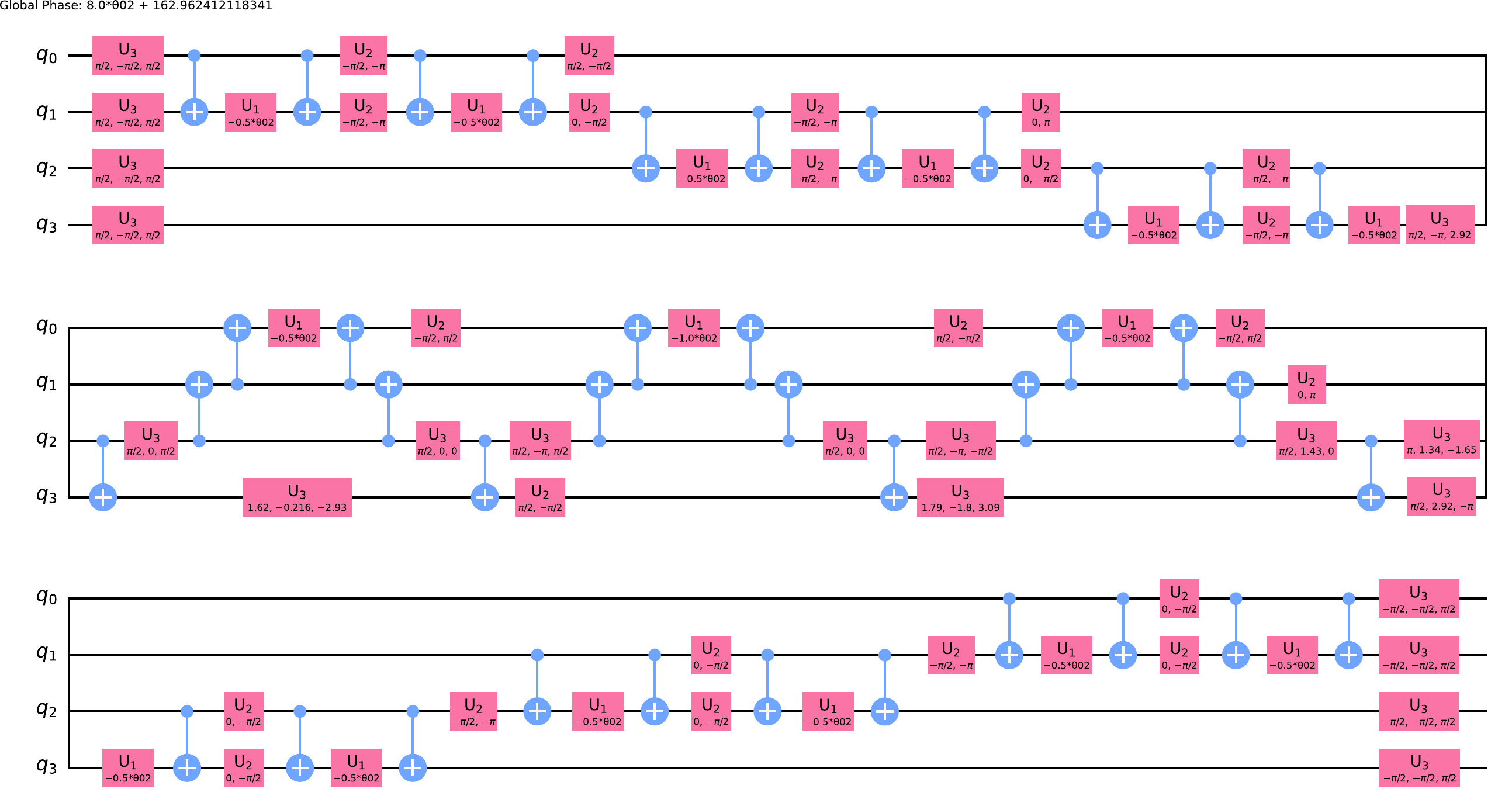}
    \caption{Nearest neighbours hopping part of Trotter circuit used in Sec.~\ref{Results.Dynamics} shown for the qubits corresponding to spins up only (the exact same structure applies to the qubits corresponding to spins down).}
    \label{fig:Trotter_t}
\end{figure}

\section{Variational Dynamics} \label{Appendix:VariationalDynamics}
This approach allows the implementation of quantum dynamics with a constant depth circuit at the cost of confining the dynamics into a given subspace of the full Hilbert space; in addition, the propagation of the time-dependent parameters requires the measurements of additional matrix elements to build the corresponding equation of motion~\cite{Yuan2019}.

\begin{figure}[h!]
    \centering
    \includegraphics[width=\linewidth]{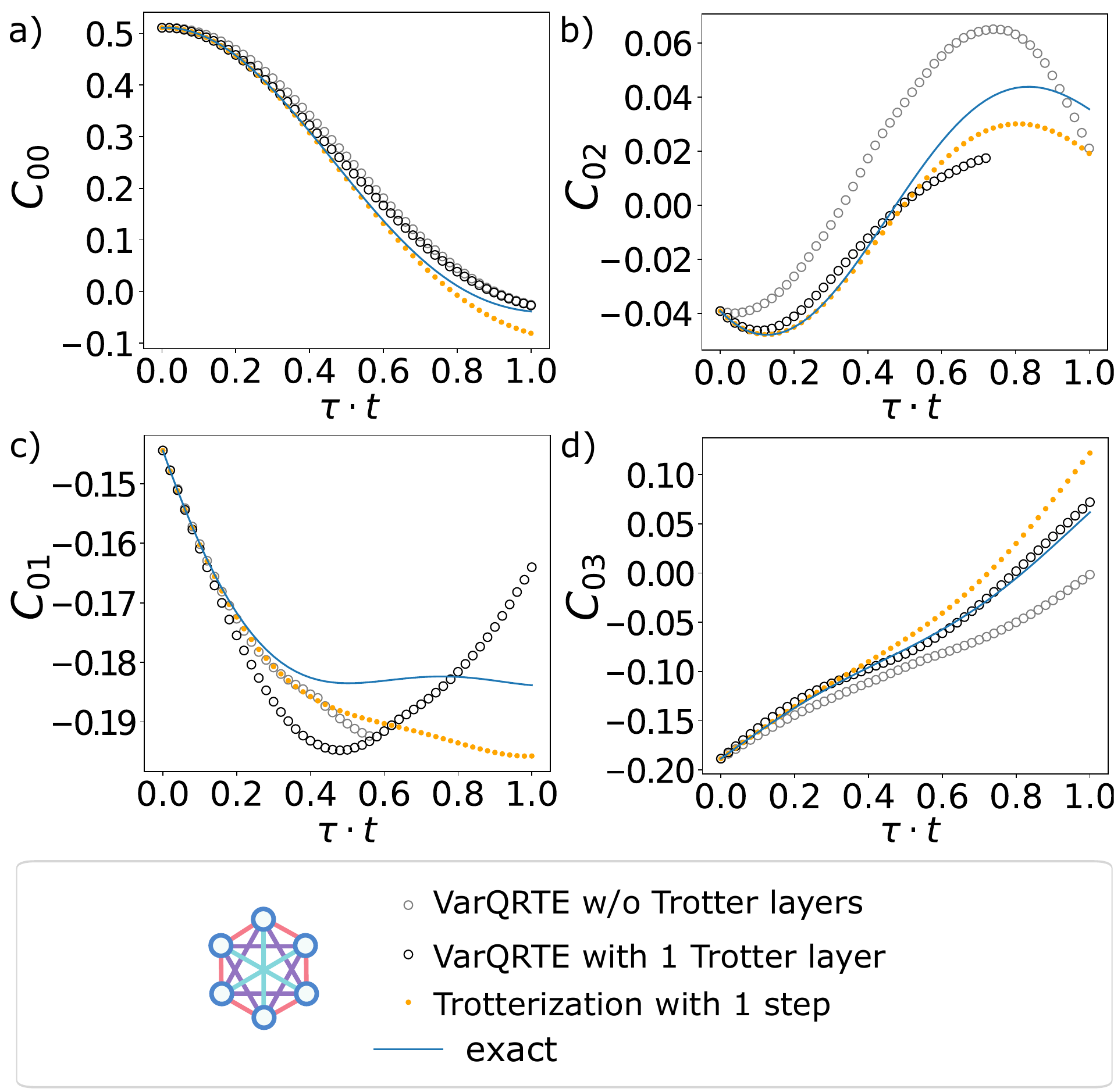}
    \caption{Dynamical spin-spin correlations $C_{0j}$ as a function of $\tau$ for the $L=6$ Hubbard model in the half-filling regime (number of particles $N=L$), with onsite interaction energy $U/t=0.5$, and next-to-nearest hopping energy $t'/t=0.5$. Full lines represent results from statevector calculations performed with \texttt{numpy}, while orange dots correspond to values obtained with the second-order Trotter formula with 1 repetition of the corresponding circuit, light grey empty circles correspond to the variational quantum real time evolution (VarQRTE) with our static ansatz, dark grey empty circles correspond to the VarQRTE with our static ansatz and an additional Trotter layer  (see legend). 
    The system was initialized in the ground state using a VQE with a convergence threshold of $5\cdot 10^{-3}$ and a circuit ansatz composed of $d=4$ repetitions.}
    \label{fig:gf_n6}
\end{figure}

Here, we provide a comparison of the results obtained with Trotter-based and variational dynamics (See Fig.~\ref{fig:gf_n6}). To this end, we compute the dynamical spin-spin correlations $C_{0j}$ (as defined in Sec.~\ref{Results.Dynamics}) 
using three approaches: (i) Trotterisation procedure (as described in Sec.~\ref{Methods.Trotterization}) with one repetition of the circuit; (ii) the variational quantum real time evolution (VarQRTE) with the same static ansatz as used for the VQE; (iii) VarQRTE with the static ansatz augmented with a repetition of each circuit in Figs.~\ref{fig:Trotter_U},~\ref{fig:Trotter_t} and~\ref{fig:Trotter_tdash} and treating the parameters ($\theta 00$, $\theta 01$, $\theta 02$) as variational. 

While the variational approach may give better results in some cases (compared to a single Trotter step dynamics with increasing $\delta t$), the choice of an adequate variational circuit that allows reaching long time scales is still hard to guess. In addition, the number of measurements needed to determine the equations of motion of the variational parameters introduces additional computational costs, which are not present in the Trotter implementation. Further studies are required to assess under which conditions the variational approach can improve over the Trotter scheme. 

\section{IBM Quantum heavy hex lattice mapping} \label{Appendix:Mapping}
To obtain general scaling bounds for mapping our model to the heavy hex lattice, we can consider the Hubbard model of arbitrary size with all-to-all hopping interactions (restricted to the same spin types) and onsite interactions (restricted to different spin types), for only a single ansatz layer. In this case, the number of hopping interactions is $\binom L2 + \binom L2 = 2\binom L2$, while the number of onsite interactions is $L$. We observe that the number of hopping interactions scales with one power more than the number of onsite interactions, so it is necessary to minimize the number of swap gates when mapping the system to the heavy hex lattice. Note that this conclusion is contingent upon the relative scaling of the numbers of ``onsite'' and ``hopping'' gates in the ansatz.

For this reason, we want to keep the spin of the same type as close as possible; one way to achieve that is using the fact that hexagons tile the plane perfectly --- we can roughly handle the spin families somewhat independently by bundling them into two hexagonal clusters --- minimizing the leading source of swap gate complexity. Then, the two clusters can be connected in some way that will of course affect the final number of necessary swap gates, however not up to a leading order. If the clusters are incomplete, we try to increase the number of their common edges as much as possible, possibly by reordering some fringe edges, again without affecting the leading complexity term, since the number of fringe edges scales in system size with at least one power less than the number of edges in the bulk of the cluster.

Now an upper bound estimate of the number of swap gates needed can be found taking a simple approach. Focusing on each spin family separately, for each particle $i$, we find the shortest path to any other particle $j$, e.g., denoted by $d_{ij}$. Then the number of swaps needed to render $j$ adjacent to $i$ is equal to $\max\{0, d_{ij} - 1\}$, and additionally the same number of swaps is needed to return $j$ to its initial position. This procedure can be repeated for all ordered pairs $(i,j)$ such that $j > i$, avoiding double counting.

Although this particular procedure of nearing different particles is suboptimal because in the process of $j$ moving towards $i$, $j$ will in general reach other particles before, it gives an upper bound on the scaling of the number of swap gates of $\mathcal O(L^3)$, as seen from two contributions. First, there are $\mathcal O(L^2)$ pairs of interacting particles. Second, the number of swaps needed to make each of them adjacent is $\mathcal O(L)$, which is an upper bound found in the case when all particles in the cluster are ordered on its fringe (evidently, the paths through the bulk in hexagonal ordering will cumulatively be shorter). In general, for an ansatz with $d$ layers, the complexity becomes $\mathcal O(dL^3)$.

Finally, we note that for particular maps described in Table~\ref{tab:Scaling} and Fig.~\ref{fig:Mapping}, there are two orders of magnitude of $L$ less in circuit complexity than in the general system we considered. One order difference is due to the fact that each particle interacts with a constant number of neighbours (compared to almost all of them), while the second is a consequence of the average distance between the neighbours being bounded by a constant (i.e., not scaling linearly with system size).

\end{document}